\documentclass[runningheads]{llncs}

\usepackage[T1]{fontenc}
\usepackage[title]{appendix}
\usepackage{xcolor}
\usepackage{url}
\usepackage{amssymb}
\usepackage[style=numeric]{biblatex}

\usepackage{multicol}

\usepackage{fancyhdr} 
\usepackage{graphicx}

\begin{document}

\title{An incremental hybrid adaptive network-based IDS in Software Defined Networks to detect stealth attacks}

\author{Abdullah H Alqahtani 
}

\institute{Department of Computer Science, The University of Sheffield, Sheffield, S10 2TN, UK
\email{ahalqahtani1@sheffield.ac.uk}\\
\url{https://www.sheffield.ac.uk}}
\maketitle              

\begin{abstract}

Network attacks have became increasingly more sophisticated and stealthy due to the advances in technologies and the growing sophistication of attackers. Advanced Persistent Threats (APTs) are a type of attack that implement a wide range of strategies to evade detection and be under the defence's radar. Software Defined Network (SDN) is a network paradigm that implements dynamic configuration by separating the control plane from the network plane. This approach improves security aspects by facilitating the employment of network intrusion detection systems.
Implementing Machine Learning (ML) techniques in Intrusion Detection Systems (IDSs) is widely used to detect such attacks but has a challenge when the data distribution changes. \emph{Concept drift} is a term that describes the change in the relationship between the input data and the target value (label or class). The model is expected to degrade as certain forms of change occur. In this paper, the primary form of change will be in user behaviour (particularly changes in attacker behaviour).  It is essential for a model to adapt itself to deviations in data distribution. SDN can help in monitoring changes in data distribution. This paper discusses changes in stealth attacker behaviour. The work described here investigates various concept drift detection algorithms. An incremental hybrid adaptive Network Intrusion Detection System (NIDS) is proposed to tackle the issue of concept drift in SDN. It can detect known and unknown attacks. The model is evaluated over different datasets showing promising results.

\end{abstract}
%
%
%
\section{Introduction}
\label{sec:Introduction}

Software-Defined Networks (SDNs) offer improved network flexibility by enabling the dynamic configuration of the network infrastructure. This is, in part, a result of the architectural choice to separate the control and data planes of the network. The centralisation of the SDN controller provides visibility over the entire network so monitoring the network and collecting data from network devices is much easier than in traditional networks \cite{kreutz2014software}.  A number of researchers have sought to use the features of SDN networks to detect network attacks \cite{abubakar2017machine, ajaeiya2017flow, girdler2021implementing, sultana2019survey}. 

However, SDNs bring with them specific threats. For example, SDNs rely on nodes maintaining flow control rules that tell a node how to handle various packets. Knowing those rules (or even some of them) can help an attacker. This gives rise to so-called flow rule reconstruction attacks. Protecting against such inference attacks remains a challenge. Attackers determined to infer such information can launch attacks that have characteristics of Advanced Persistent Threats (APTs), typically seeking to achieve their goals stealthily, and following a strategy of keeping under any detection thresholds by employing slow movement and generating or engaging in only a low volume of traffic. Rather than engage in a ‘smash-and-grab’ as would traditional attackers, APTs might stay days, weeks, months or years in the victim's network and seek to achieve their goals while minimising the risk of raising suspicion \cite{alshamrani2019survey, halabi2021protecting}.

Nowadays, machine learning (ML) techniques are widely employed by Network Intrusion Detection Systems (NIDSs) to detect malicious activities inside the network. In NIDS, known attacks can be detected by \emph{signature-based} detection techniques. Such techniques detect predefined patterns of malware or malicious behaviour.  If an attack is unknown, as with most APTs, \emph{anomaly-based} detection techniques give better chances for detection. In an anomaly-based detection technique, the model profiles the normal user's behaviour by training using only normal traffic (or other behaviour). When an attack happens, malicious activities are detected as deviations from the pattern characteristics of profiled normal behaviour.

One of the main challenges in ML-based intrusion detection systems is that the environment from which data is sampled changes over time. That is the distribution of the data to be classified changes. Such change may occur, for example, as a result of a change in the behaviour of both benign users and attackers. 
A change in the network environment, e.g. adding more users or devices to the network, is another reason causing a change in data distribution. \emph{Concept drift} is the change over time in the relationship between the input data and the predicted output (i.e., how inputs are classified).  If the drift is not accounted for, the results of a predictive ML model will deteriorate \cite{lu2018learning}. 

When concept drift occurs, there is a need for the IDS to adapt itself to change by training over new data; it cannot simply depend on old data. This paper combines two incremental adaptive techniques to improve the system's performance. An Adaptive Random Forests (ARF) (a signature-based detection technique) and an Adaptive One-Class SVM (an anomaly-based detection technique) are proposed to work in a hybrid approach to detect known and unknown attacks respectively. Combining these signature-based and anomaly-based approaches improves detection and outperforms existing standard ML techniques. Various experiments are conducted over different datasets to evaluate the proposed model. The adoption of drift detection and response strategies counters model degradation due to concept drift. The proposed model is shown to detect known and unknown stealth attacks in SDN as well as traditional attacks.

\section{Background}
\subsection{Concept Drift}
\label{ch2: ConceptDrift}

\begin{figure}
  \includegraphics[width=\linewidth]{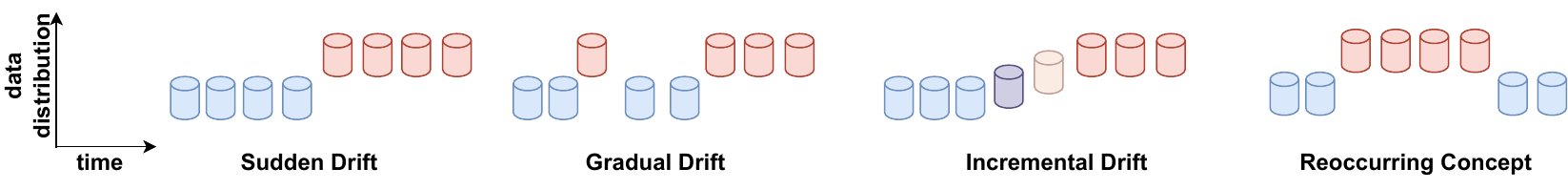}
  \caption{Concept drift types (adapted from \cite{lu2018learning})}
  \label{fig: CDtype}
\end{figure}

The distribution of data an intrusion detector is presented with may change for a number of reasons, e.g. updating or replacing network devices, increasing or decreasing the number of users or devices in the network, and user behaviour changes. A classifier's performance may degrade if the data witnessed under training no longer adequately represents the current situation. The term \emph{concept drift} is often used to describe this. Concept drift can be understood as changes in the relationship between the input and the target outputs of a classifier \cite{lu2018learning}. It may be sudden/abrupt, incremental, gradual, or reoccurring (see Figure \ref{fig: CDtype}). There are three main categories of concept drift detection algorithms \cite{lu2018learning}:
\begin{itemize}
    \item \textbf{Error rate-based drift detection:} This technique calculates the error rates raised by the base classifier. A concept drift happens if the error rate is increased or decreased dramatically. The most popular error rate-based drift detection algorithms are ADaptive WINdowing (ADWIN) \cite{bifet2007learning}, Drift Detection Method (DDM) \cite{gama2004learning}, Early Drift Detection Method (EDDM) \cite{baena2006early}, Hoeffding's bounds (HDDM\_A and HDDM\_W) \cite{frias2014online}, and Kolmogorov-Smirnov Windowing (KSWIN) \cite{raab2020reactive}.
    \item  \textbf{Data distribution-based drift detection:} These algorithms use distance functions to calculate the pattern changes between old and new data. They have the advantage of reporting the time and location of the drift. However, they have high computational costs \cite{lu2018learning} \cite{lu2016concept}. Examples of these techniques are kdq-tree \cite{dasu2006information}, CM \cite{lu2016concept} \cite{lu2014concept}, RD \cite{kifer2004detecting}, SCD \cite{song2007statistical}, EDE \cite{gu2016concept}, SyncStream \cite{shao2014prototype}, PCA-CD \cite{qahtan2015pca}, LSDD-CDT \cite{bu2016pdf}, LSDD-INC \cite{bu2017incremental}, and LDD-DSDA \cite{liu2017regional}.
    \item \textbf{Multiple hypothesis test drift detection:} These approaches apply multiple techniques (similar to those in the previous two categories). They apply different ways to detect the concept drift, but in general, they calculate the drift by employing different hypothesis tests in a parallel or hierarchical approach \cite{lu2018learning}.
\end{itemize}

\subsection{Incremental learning model}
\begin{figure}
  \includegraphics[width=\linewidth]{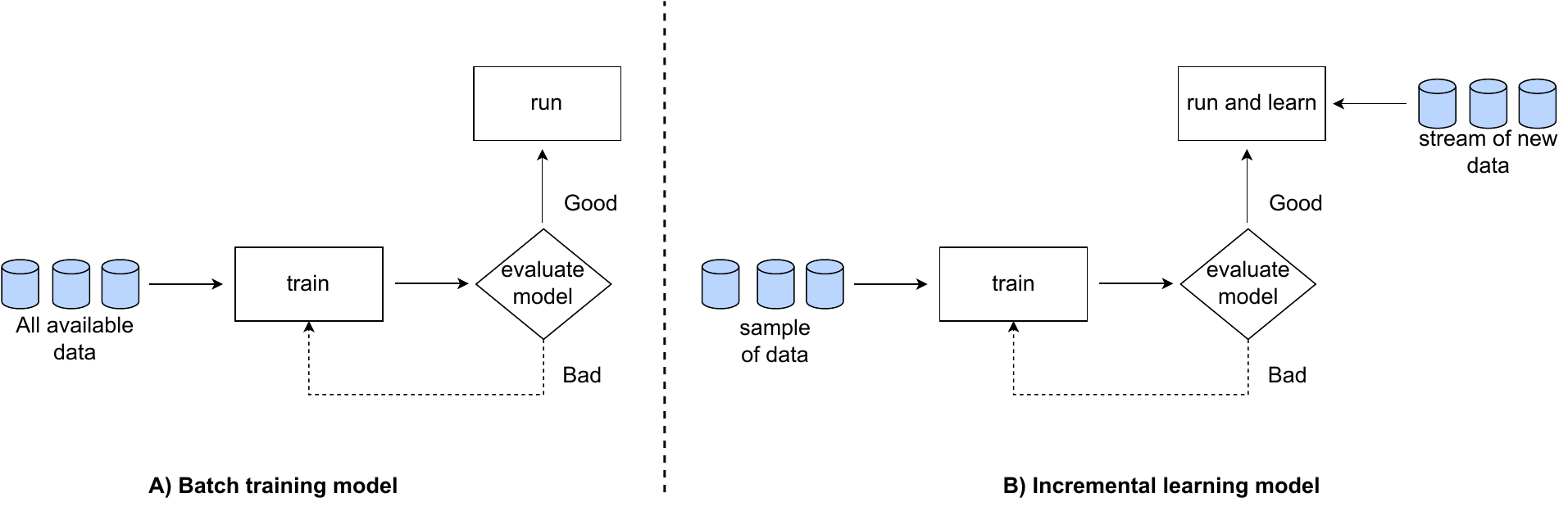}
  \caption{Difference between batch training and incremental learning model}
  \label{fig: batchVSincremental}
\end{figure}

To overcome the issue raised by concept drift (changes in data distribution), the model should learn incrementally over new instances. 
In standard machine learning techniques, the model is typically trained offline, with all training data available at the time of training the model. The model can be deployed when the training is completed. It can be trained later using a new batch of data. This type of training is called batch (or offline) training. The incremental learning model does not require the entire data to be available initially. Instead, the data is processed sequentially, and the model updates itself as new data arrives \cite{gama2014survey}. Figure \ref{fig: batchVSincremental} illustrates the difference between the two models.

Our model uses two incremental learning techniques to adapt to concept drift:

\begin{itemize}
    \item Adaptive Random Forest (ARF): A Random Forest is an ensemble ML technique that consists of a large number of individual decision trees (DTs). The target class is the class most voted for by these trees \cite{breiman2001random}. An Adaptive Random Forest (ARF) is an optimised technique for Random Forest that enables it to classify evolving data streams. It employs resampling methods and adaptive operators to overcome the issues of different types of concept drift. 
    ARF implements a drift detector that observes the accuracy of the prediction on every incoming data item. Every tree has a threshold that indicates the change is a warning or a drift. When a tree is identified as suffering from concept drift a warning is triggered and a new tree is trained in the background to replace it  \cite{gomes2017adaptive}.
    \item One-class Support Vector Machine (OC-SVM): An OC-SVM is an unsupervised learning technique that can be used for anomaly-based detection. It learns to draw boundaries between normal data and anomalies. In the training phase, the model is trained solely over normal instances. In the testing (or production) phase, when a new instance deviates from the normal data, it is considered an anomaly. OC-SVM is a batch learning model which cannot work on streaming data. To overcome this issue, Stochastic Gradient Descent (SGD) \cite{robbins1951stochastic} is augmented to be used with One-Class SVM as an online version for OC-SVM (giving rise to an Adaptive One-Class SVM) to detect anomalies in streaming data. SGD computes the gradient when a single instance (or mini-batch) of data arrives. Another advantage of  SGD is its highly efficient for working on large data \cite{OnLineSGDOneClassSVM}. 
\end{itemize}

\subsection{Paper organisation}
The rest of this paper is organised as follows. Section \ref{RelatedWorks} summarises the capabilities of existing proposed approaches. Section \ref{Contributions} shows our contributions. An explanation of the proposed model is given in section \ref{ProposedScheme}. Evaluation results are presented in section \ref{results}. Conclusions and future work are given in section \ref{conclusion}.

\section{Related Work}
\label{RelatedWorks}

Adaptive approaches have been brought to bear for intrusion detection in non-SDN environments. An ensemble classifier has been proposed to detect concept drift caused by the existence of new intrusions \cite{yuan2018concept}. \emph{HDDM} \cite{frias2014online} is used as the concept drift detector and the system scores 0.94 in accuracy. Another ensemble classifier \cite{martindale2020ensemble} uses Hoeffding Adaptive Tree and Adaptive Random Forest to detect network attacks and uses ADWIN for concept drift. A network-based IDS uses a Page-Hinkley test (PHT) to detect concept drift to improve a deep learning classifier \cite{andresini2021network}. The classifier has high accuracy (0.99) and is applied on the CICIDS2017 \cite{cicids2017} dataset (which contains only traditional attacks). The model is not generalised as it is trained over the dataset without removing features that could cause over-fitting. A stream learning IDS \cite{horchulhack2022stream} is proposed for detecting concept drift. CIDD-ADODNN \cite{priya2021deep} uses Adadelta optimiser-based deep neural networks (ADODNN) to classify imbalanced data. The proposal used anomaly-based detection and was evaluated over three different datasets.  Optimised Adaptive and Sliding Windowing (OASW) \cite{yang2021lightweight} is a window-based drift detection technique proposed for Internet of Things systems. The authors proposed an optimisation for LightGBM to detect anomalies in such systems. It is evaluated over IoTID20 dataset \cite{ullah2020scheme} and NSL-KDD \cite{nsldataset}. The accuracy results are 0.99 on IoTID20 \cite{ullah2020scheme} and 0.98 over the NSL-KDD dataset. These schemes are not proposed to detect stealth attacks or work on SDN. Table \ref{table: RelatedWorksCD} shows a comparison between the aforementioned studies and our proposed model. 

\begin{table*}[h]
\caption{Related works Comparison}
\begin{center}
\setlength\tabcolsep{3 pt}
\begin{tabular}{|c | c |c |c| c| p{4 cm}|  } 
\hline 
\textbf{Scheme} &  \textbf{SDN} & \textbf{Stealth}& \textbf{Accuracy} & \textbf{Technique}& \textbf{Dataset}\\ [0.5ex] 
\hline

\cite{yuan2018concept} &  $\times$ & $\times$& 0.94 & HDDM & KDD Cup 99 \cite{kdd}\\
\hline

\cite{andresini2021network} &  $\times$ & $\times$& 0.99    & PHT & CICIDS2017  \\
\hline

\cite{horchulhack2022stream} &  $\times$ & $\times$& 0.96   & bespoke & Fine-grained Intrusion Dataset (FGD) \cite{dos2021improving} \\
\hline

\cite{priya2021deep} &  $\times$ & $\times$ & 0.95  & ADWIN & NSL-KDD \\
                     &           &          &0.93  &   & Spam \cite{SpamDS}\\
                     &           &          &0.76   &   & Chess\cite{chessDS} \\
\hline

\cite{martindale2020ensemble}  &  $\times$ & $\times$& -  & ADWIN & KDD Cup 99 \\
\hline

\cite{yang2021lightweight}  &  $\times$ & $\times$& 0.99  & ADWIN & IoTID20   \\
                            &           &          & 0.98   &       & NSL-KDD \\
\hline

Proposed model & $\checkmark$ & $\checkmark$ & 0.99   & ADWIN & APT-SDNdataset \cite{APTSDNdataset} \\
               &              &              & 0.98   &       & DAPT 2020 \cite{myneni2020dapt}\\
               &              &              & 0.99   &       &  CICIDS 2017 \cite{cicids2017}\\
               &              &              & 0.94   &       & InSDN \cite{elsayed2020insdn}\\
\hline

\end{tabular}

\end{center}
\label{table: RelatedWorksCD}
\end{table*}

\section{Contributions}
\label{Contributions}

The review of section \ref{RelatedWorks} reveals a lack of work dealing with concept drift in SDN, especially when attacks are stealthy. Our hybrid model in \cite{alqahtanidetecting} can detect such attacks in a non-streaming environment. In this paper, we propose to detect such attacks in a streaming environment.

Our focus on adaptivity is driven in part by the fact that APTs may engineer concept drift as a stealth strategy. They may continually adapt their behaviour so that they maintain sufficient similarities with normal behaviours to remain undetected \cite{chandola2009anomaly}. They may stay dormant for a period of time or employ sleep functions between their activities \cite{alqahtanidetecting}. Moreover, the majority of these attacks go through multiple stages and at every stage, they have one or more tools that may create new classes of attack. New patterns may emerge as a result of these changes \cite{halabi2021protecting} \cite{alshamrani2019survey} \cite{mulimani2021adaptive}. It is essential to retrain the ML-based NIDS to track the concept drift and make the model more adaptive to any changes in data distribution. In our experiments, we use various datasets to evaluate the proposed model (when attacker changes their behaviour).

The contributions of this paper are:
\begin{itemize}
   
    \item The first proposal of incremental adaptive hybrid IDS to overcome concept drift caused by stealth attackers in SDN.
    \item A comparison between the proposed hybrid model (adopting concept drift detection) and standard hybrid model without implementing concept drift detection techniques.
     \item Evaluation of the most popular concept drift detection techniques.
    \item Evaluation of the proposed system over different datasets, reflecting various attacks and scenarios (including known and unknown attacks).
    \item Determination of optimal parameters using hyper-parameter tuning.
\end{itemize}

\section{Proposed Scheme}
\label{ProposedScheme}

Given a period of time [0,$t$], streaming data arrives as a sequence of packets $K_{0,t} = \{s_{0}, s_{1}, s_{2},..., s_{n}\}$, where $s_{i} =(X_{i},y_{i})$ is one sample. $X_{i}$ is a vector of independent features which can be labelled appropriately with  $y_{i} \in \{0,1\}$. Let $P_{t}(X,y)$ represent the \emph{joint distribution} of $X$ and $y$ at time $t$. (It is a probability density function.) Concept drift happens when $P_{t}(X,y)$  changes over time. Thus, it occurs between two time points $t_{0}$ and $t_{1}$  if  $\exists X \cdot P_{t0}(X,y) \neq P_{t1}(X,y)$ \cite{gama2014survey}. This is defined in terms of the \emph{data} sampled. For example, if a new attack is developed after time $t_{0}$ and launched, corresponding to   $X'$ say, then  $P_{t0}(X',1)=0$ (the attack never occurs in the data and so has zero density) but $P_{t1}(X',1)\neq 0$ (the attack does occur and so has non-zero density).   Technically, a specific vector of features $X$ could in some circumstances correspond to malicious action but be benign in others, though often the assignation will be unequivocal. ML-based models that seek to predict $y$ from $X$ must adapt to ensure continued high performance.

\begin{figure} 
\centering
  \includegraphics[width=0.65 \textwidth]{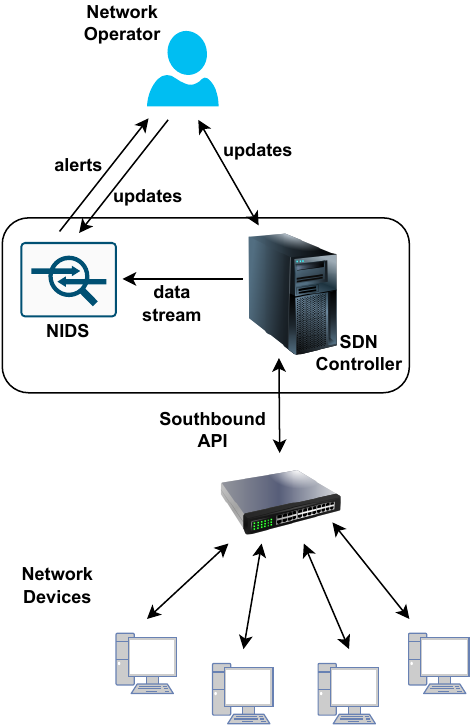}
  \caption{The proposed NIDS model architecture}
\label{fig: SysArch}
\end{figure}

\begin{figure*} 
  \includegraphics[width=\linewidth]{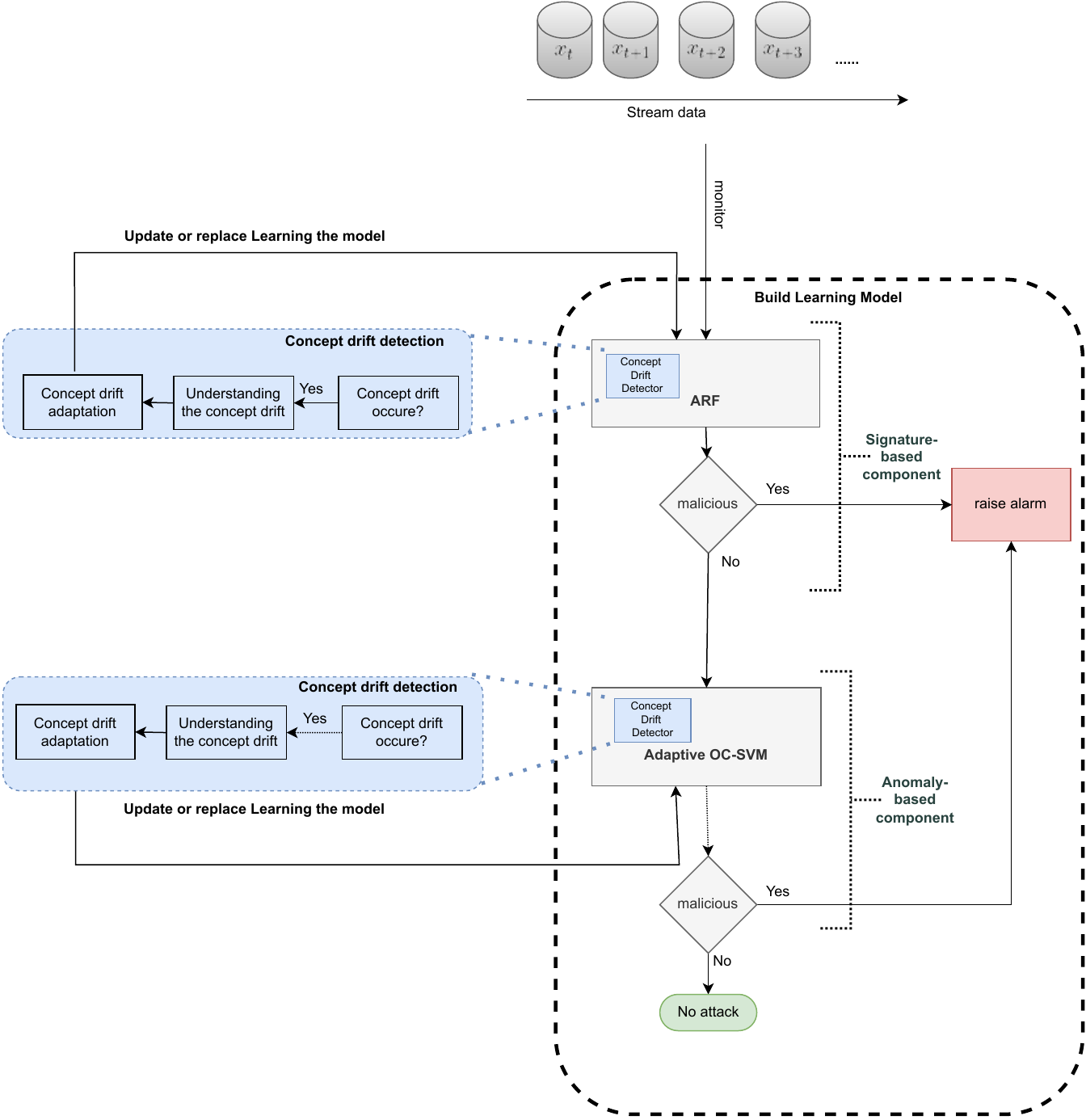}
  \caption{The proposed adaptive hybrid model}
\label{fig: AdaptiveModel}
\end{figure*}

The proposed hybrid system combines signature-based and anomaly-based detection techniques using Adaptive Random Forest (ARF) and adaptive One-Class SVM. The model monitors networks through the SDN controller. Figure \ref{fig: SysArch} shows the place of the proposed NIDS. The details of the detection of the model is illustrated in Figure \ref{fig: AdaptiveModel}. Every instance that arrives at the system is, firstly, examined by the signature-based ARF module (discussed in section  \ref{SigModel}). If the module classifies that instance as a known attack, then there is no need for further investigation by the anomaly-based detection module. If the packet is classified as benign by the signature-based module, it is passed to the anomaly-based detection module (Adaptive One-class SVM) (discussed in section  \ref{AnomalyModel}) to check if it is an unknown attack. The administrator is notified if the signature or anomaly-based module flags an instance as an attack. The concept drift is detected using ADWIN, a concept drift detector, in both classifiers, and the model is updated. The model can detect known and unknown attacks. Both are incremental and can adapt themselves against concept drifts.

\subsection{Signature-based detection module}
\label{SigModel}
The first detection phase is the signature-based detection module, where the classifier is trained over labelled data to learn the pattern of malicious and normal activities. Adaptive Random Forest (ARF) is employed as the incremental adaptive technique.
 The ARF uses a drift detector in every tree, which monitors drifts and warnings. In the case of a warning, the trees in the background will train new trees. If a drift is detected, the primary model will be replaced by the trained trees. Hyper-parameter tuning is applied to ensure the model is generalised (avoiding overfitting or underfitting). Table \ref{table:HyperParameters} shows the selected parameters of ARF.

\subsection{Anomaly-based detection module}
\label{AnomalyModel}
Adaptive One-class SVM is an anomaly detection algorithm. The classifier is trained over the normal data and tested over both normal and malicious samples. During the training, it develops a profile of normal traffic. If the incoming traffic differs from the normal data that it is trained on, it is considered an attack. It also adapts itself incrementally when the data distribution is changed. The new data points are compared to the existing model. A change in the distribution of the data points indicates a concept drift is happening. SVMs and their variants have been shown to be highly effective classifiers across many domains. The selected parameters after applying hyper-parameter tuning are shown in Table \ref{table:HyperParameters}.

\begin{table}
\caption{Selected parameters of ARF and OC-SVM}
\centering
\begin{center}
\begin{tabular}{|c | p{4.8 cm} | c|} 
\hline \textbf{parameter} & \textbf{description} & \textbf{optimal value} \\ [0.5ex] 
\hline\hline
\multicolumn{3}{|l|}{\textbf{AdaptiveRandomForestClassifier (ARF)}} \\
\hline
max\_features & Maximum number of features in every single run & 5  \\
\hline
n\_estimators  & Number of trees  & 7\\ 
\hline
drift\_detection\_method & concept drift detection technique  & ADWIN\\ 
\hline
\multicolumn{3}{|l|}{\textbf{Adaptive One-Class SVM (OC-SVM)}} \\
\hline
$\nu$ & This controls the fraction of outliers in the system & 0.2 \\
\hline
Gamma & Kernel coefficient & 0.9\\
\hline

\end{tabular}
\end{center}
\label{table:HyperParameters}%
\end{table}

\section{Evaluation and Results}
\label{results}
APT campaigns go through multiple stages and different types of attacks are involved in every stage. Moreover, there is no one pattern for APT attacks \cite{alshamrani2019survey}. In this paper, a number of experiments were conducted on different datasets to show the effectiveness of the proposed model in detecting attacks and dealing with concept drift (when attacker changes their behaviour). These datasets include different attacks and scenarios that present a change in attacker behaviour. Table \ref{table:DSsummary} presents a summary of every dataset and the aim of conducting the experiment over that dataset.
In all experiments, various evaluation standard measures are recorded: \emph{Accuracy, Recall, Precision, and F1-score}.

\begin{table}
\caption{A summary of datasets and experiments objectives}
\begin{center}
\begin{tabular}{| c |p{4.9 cm} |p{4.9 cm} |} 
\hline 
\textbf{Dataset} & \textbf{Dataset feature} & \textbf{aims of the experiment} \\ [0.5ex] 
\hline
DAPT 2020 \cite{myneni2020dapt} & Contains APT traffic and a change in attacker behaviour (concept drift) & - evaluate the model in detecting APTs \\
          &                                                                         & - evaluate the model against concept drift \\
          &                                                                         & - compare the standard ML techniques with those after adaptation \\
\hline
APT-SDNdatase \cite{APTSDNdataset} & stealth attacks on SDN (reconstructing flow rules in SDN) & evaluate the model on detecting stealth attacks in SDN\\
\hline
InSDN \cite{elsayed2020insdn} & Traditional attacks in SDN over different days & - evaluate the model against various attacks in SDN \\
          &                                           & - evaluate the model against concept drift over SDN network \\
\hline
CICIS 2017 \cite{cicids2017}& A benchmark IDS dataset has various attacks over different days & - evaluate the model against different traditional attacks\\
          &                                                                   & - evaluate the model against concept drift\\
\hline

\end{tabular}

\end{center}
\label{table:DSsummary}
\end{table}

\begin{enumerate}

\item{\textbf{Model evaluation over APT dataset:}}
\label{Experiment3}
DAPT 2020 \cite{myneni2020dapt} is an APT-based dataset used in this experiment to evaluate the proposed model against APT attacks and attackers' behaviour changes (concept drift). The approach of conducting normal traffic and various attacks over different days makes this dataset very useful for evaluation. In Appendix \ref{appendexA}, Table \ref{table: DAPT2020discription} presents details of the attack and normal traffic scenario in the dataset. The dataset is prepared by removing columns that can affect the system's performance. Flow ID, Source IP, Source Port, Destination IP, Destination Port and Timestamp are the features that are removed from the dataset. Null values are replaced by zeros, and labels are converted to (0) and (1) as the system is a binary classifier. The dataset's features are scaled using scikit-learn's StandardScaler \cite{StandardScaler}.  The first experiment compares error rate-based drift detection techniques: ADWIN, DDM, EDDM, HDDM\_A, HDDM\_W, PageHinkle, and KSWIN. The system gives significant results for detecting APTs, scoring 0.98 in Accuracy, Recall, Precision and F1 using any of these techniques (concept drift detector). Table \ref{tab:DAPTresultsErrorBased} shows the results of all techniques with a negligible variation in scores.

\begin{table}
\caption{Results for error rate-based drift detection (over DAPT 2020 dataset)}
\begin{center}
\begin{tabular}{|c | c |c |c| c|} 
\hline 
\textbf{Tech} & \textbf{Accuracy} & \textbf{Recall} & \textbf{Precision} &\textbf{F1}\\ [0.5ex] 
\hline
ADWIN & 0.9871 & 0.9882 & 0.9856 & 0.9869 \\
\hline
DDM & 0.9872 & 0.98871 & 0.9853 & 0.9870  \\
\hline
EDDM & 0.9862 & 0.9879 & 0.9839 & 0.9859 \\
\hline
HDDM\_A & 0.9867 & 0.9888 & 0.9842 & 0.9865  \\
\hline
HDDM\_W & 0.9838 & 0.9852 & 0.9818 & 0.9835  \\
\hline
PageHinkle & 0.9855 & 0.9879 & 0.9826 & 0.9852  \\
\hline
KSWIN &  0.9856 & 0.9872 & 0.9834 & 0.9853  \\
\hline
\end{tabular}

\end{center}
\label{tab:DAPTresultsErrorBased}
\end{table}

Another experiment is conducted to evaluate the implementation of the distribution-based detection technique (using kdq-tree). This achieves 0.83, 0.80, 0.84 and 0.82 in Accuracy, Recall, Precision and F1 respectively. The results are not significant, but the technique has advantages as it is unsupervised.

ADWIN will be used on the model as the concept drift detection technique. Because most error-based concept drift techniques are very similar but ADWIN has more flexibility than others. The window size (in ADWIN) is variable as it can keep expanding and shrinking, making the sliding window mechanism more flexible than other techniques \cite{bifet2007learning} \cite{lu2018learning}. In addition, it gives better results than the distribution-based concept drift algorithm (e.i. kdq-tree).

To evaluate to which extent adopting concept drift detection can improve the system, a further experiment is conducted on standard classifiers (batch learning). The standard classifiers Random Forest and One-Class SVM are employed to form the hybrid system. The results are 0.72 Accuracy, 0.72 Recall, 0.78 Precision and 0.75 in F1. Figure \ref{fig: DAPTcompare} shows how the system is improved after implementing the adaptivity (using ADWIN for error rate-based drift detection and kdq-tree for data distribution based).
 
\begin{figure} [t!]
  \includegraphics[width=\linewidth]{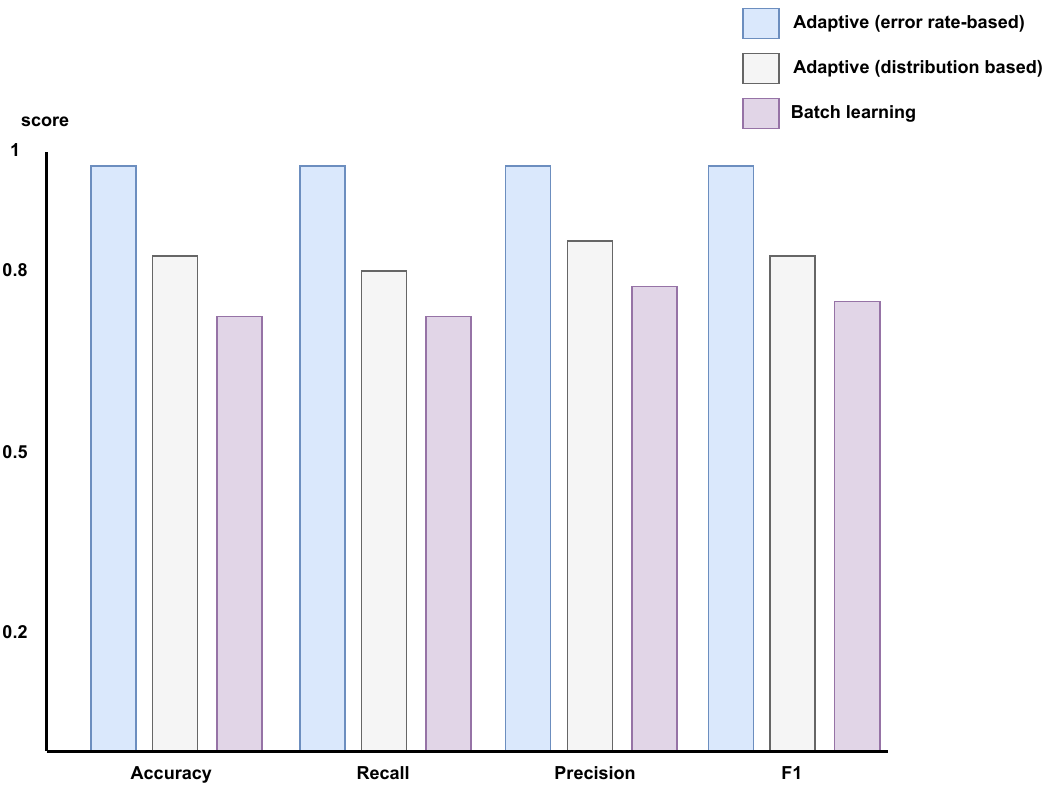}
  \caption{A comparison on the evaluation of batch learning and the proposed incremental model over the DAPT 2020 dataset}
  \label{fig: DAPTcompare}
\end{figure}

\item{\textbf{Model evaluation over stealth SDN-based dataset:}}
\label{Experiments 1}
These experiments aim to evaluate how well the proposed hybrid model detects stealth attacks over SDN network. APT-SDNdataset \cite{APTSDNdataset} is the dataset used in the experiments. The dataset has a probe scan that reconstructs flow rules in a stealthy manner \cite{alqahtani2022enhanced}. The first experiment uses ADWIN as the concept drift detector technique. It achieves with 0.99 in accuracy and recall. 0.95 and 0.97 in precision and F1 respectively.

The second experiment uses the data distribution-based drift detection technique. The kdq-tree scores 0.99, 0.96, 0.96 and 0.96 in accuracy, recall, precision and F1 respectively. The results show that the error-based detection algorithm (ADWIN) performs better than kdq-tree. Precision (which affects F1) is much lower in kdq-tree, compared to ADWIN results, due to the high percentage of false positives.

\item{\textbf{Model evaluation over SDN-based dataset:}}
\label{Experiment2}
This experiment is conducted over an SDN-based dataset to show how the model can detect various attacks over SDN networks and adapt the model over the concept drift issue. InSDN \cite{elsayed2020insdn} is a dataset made up from SDN-based traffic. It is divided into three groups based on the type of traffic and the targeted machine. The \emph{ normal group}: contains just normal traffic with no attacks; the \emph{ Metasploitable-2 group}: includes attacks that targeted the Metasploitable-2 machine; and the \emph{OVS group}: contains attacks that target the OVS switch. The attacks in the dataset are DoS, DDoS, web attacks(XSS and SQL injection), R2L, Malware(botnet), probe and U2R (exploitation) attacks. In Appendix \ref{appendexB}, Table \ref{table:InSDN} gives a summary of attack details and the normal user behaviour. More details about the dataset are in \cite{elsayed2020insdn}.
Firstly, the model is evaluated using an error-based concept drift detection technique. ADWIN is used in the experiment scoring 0.94 in accuracy, 1 in recall, 0.93 in precision and 0.96 in F1. Another experiment is conducted using the distribution-based technique. kdq-tree is the algorithm used in the experiment, and the results are 0.94, 0.99, 0.93 and 0.96 in accuracy, recall, precision and F1, respectively. The results show significant performance in detecting actual attacks correctly (True Positives). But this came with a price, as it classified some normal instances as attacks (False Positives).

\item{\textbf{Model evaluation over traditional attacks:}}
\label{Experiment4}
A benchmark and recent dataset in the intrusion detection systems is used in this experiment to evaluate the model against traditional attacks and concept drift problems. We use the CICIDS 2017 dataset in the experiment. It has a large amount of data recorded over five days and contains different types of attacks. In Appendix \ref{appendexC}, Table \ref{table:CICIDS2017} shows the normal and attack details. The first day is just normal traffic, but the following days have different attacks every day. This is a suitable scenario for concept drift as the attacker changes their behaviour over time. 
In this work, the dataset is prepared by removing all features that may expose the identity of the sender/receiver or cause over-fitting to the system. The removed features are Flow ID, Source IP, Source Port, Destination IP, Destination Port and Timestamp. The feature importance technique is applied using Random Forest \cite{RFfeature}. Just seven features, the most important of which are shown in Figure \ref{fig: CICIDSfeatureImportance}, were selected and applied. The proposed system is a binary classifier. Thus all attacks are converted to one class (1), and all normal records are assigned to (0). In the first experiment, ADWIN is used as the concept drift detector. The second experiment uses the concept drift distribution-based detector (kdq-tree). Significant results are shown in both experiments. All metrics score 0.99 while implementing ADWIN. With kdq-tree, it has very high scores in accuracy (0.99) and recall (1); however, the result in precision is 0.86 and in F1 is 0.92.

\begin{figure}[t!]
  \includegraphics[width=\linewidth]{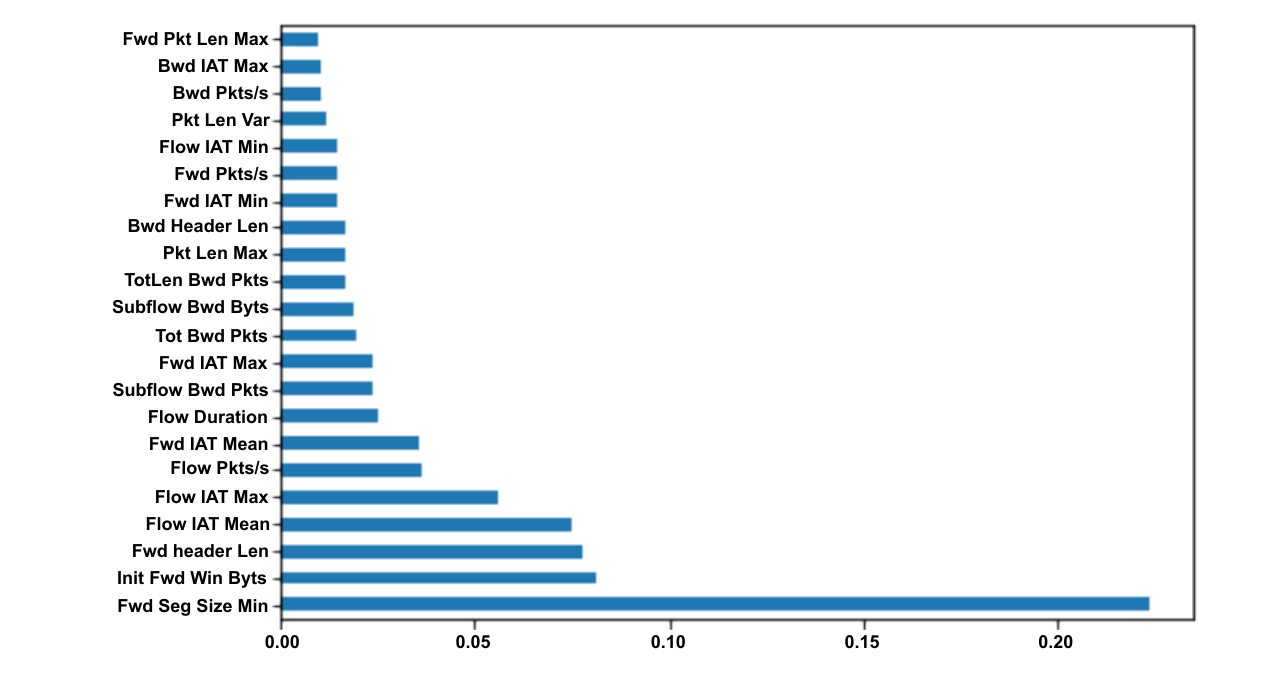}
  \caption{CICIDS 2017 feature importance}
  \label{fig: CICIDSfeatureImportance}
\end{figure}

\end{enumerate}

Table \ref{table: allExperimentsResults} presents a summary of the results of all experiments applying ADWIN and kdq-tree on different datasets.


\begin{table}
\caption{Proposed model results over different datasets}
\begin{center}
\begin{tabular}{|c | c |c |c |c| c|} 
\hline 
\textbf{Technique} & \textbf{Dataset} & \textbf{Accuracy} & \textbf{Recall} & \textbf{Precision} &\textbf{F1} \\ [0.5ex] 
\hline
ADWIN & APT-SDNdataset & 0.99 & 0.99 & 0.95 & 0.97 \\
kdq-Tree & APT-SDNdataset & 0.99 & 0.96 & 0.96 & 0.96  \\
ADWIN & InSDN & 0.94 & 1 & 0.93 & 0.96 \\
kdq-Tree & InSDN & 0.94 & 0.99 & 0.93 & 0.96  \\
ADWIN & CICIDS 2017& 0.99 & 0.99 & 0.99 & 0.99  \\
kdq-Tree & CICIDS 2017 & 0.99 & 1 & 0.86 & 0.92  \\
ADWIN & DAPT 2020 & 0.98 & 0.98 & 0.98 & 0.98  \\
kdq-Tree & DAPT 2020 & 0.83 & 0.80 & 0.84 & 0.82  \\
\hline

\end{tabular}
\end{center}
\label{table: allExperimentsResults}
\end{table}

\section{Conclusion and future works}
\label{conclusion}
Machine learning model performance will typically degrade when the data distribution is changed over time. 
APTs usually change their behaviour, causing a change in data distribution. An incremental adaptive hybrid intrusion detection system is proposed in this work. The signature-based detection (implemented by Adaptive Random Forest) and anomaly-based detection model (implemented using Adaptive One-Class SVM) are combined to detect anomalies. It can detect known and unknown stealth attacks. The system adapts itself incrementally to changes in data distribution (concept drift). ADWIN is adopted in the proposed model to detect concept drift and adapt the detection models.

The diversity of datasets used in the evaluation and the high scores obtained show how the proposed model can detect known and unseen attacks and adapt against concept drift. We recommend adapting security solutions (e.g. intrusion detection systems) incrementally. This can maintain the performance of the system even when a new attack has emerged or the attacker changes their strategy in the attack. SDN helps in this approach by making gathering and monitoring data easier.

Our proposed model can deal with concept drift that may occur based on common system evolution. Examples of these changes are producing new services, installing or removing a number of network devices or a change in the number of users. However, our proposed model is not evaluated over this scenario. In the future, we plan to evaluate the proposed model over these scenarios. Another plan is to evaluate the system using a fuller range of the most popular distribution-based concept drift techniques (e.g. CM \cite{lu2016concept} \cite{lu2014concept}, RD \cite{kifer2004detecting}, SCD \cite{song2007statistical}, EDE \cite{gu2016concept}, SyncStream \cite{shao2014prototype}, PCA-CD \cite{qahtan2015pca}, LSDD-CDT \cite{bu2016pdf}, LSDD-INC \cite{bu2017incremental}, and LDD-DSDA \cite{liu2017regional}).

\printbibliography

@article{kreutz2014software,
  title={Software-defined networking: A comprehensive survey},
  author={Kreutz, Diego and Ramos, Fernando MV and Verissimo, Paulo Esteves and Rothenberg, Christian Esteve and Azodolmolky, Siamak and Uhlig, Steve},
  journal={Proceedings of the IEEE},
  volume={103},
  number={1},
  pages={14--76},
  year={2014},
  publisher={Ieee}
}

@misc{APTSDNdataset,
  title = {APT-SDNdataset},
  url={https://github.com/APT-SDNdataset},
  journal = {GitHub},
}

@misc{chessDS,
  title = {Combining similarity in time and space for training set formation under concept drift},
  url={https://sites.google.com/site/zliobaite/resources-1},
  journal = {Intelligent Data Analysis},
}

@misc{SpamDS,
    key         = {SpamDS},
    title       = {SPAM E-mail Database},
    note        = {https://cse.usf.edu/~lohall/dm/UCIarff/spambase.arff},
    year        = 2023
}

@misc{kdd,
title={KDD Cup 1999 Data}, 
url={http://kdd.ics.uci.edu/databases/kddcup99/kddcup99.html}, 
journal={KDD Cup 1999 Data}
}

@misc{nsldataset, title={NSL-KDD dataset}, url={http://www.unb.ca/cic/datasets/nsl.html}, journal={NSL-KDD dataset}}

@misc{cicids2017, title={ntrusion Detection Evaluation Dataset (CIC-IDS2017)}, url={https://www.unb.ca/cic/datasets/ids-2017.html}, journal={UNB dataset}}

@inproceedings{ajaeiya2017flow,
  title={Flow-based intrusion detection system for SDN},
  author={Ajaeiya, Georgi A and Adalian, Nareg and Elhajj, Imad H and Kayssi, Ayman and Chehab, Ali},
  booktitle={2017 IEEE Symposium on Computers and Communications (ISCC)},
  pages={787--793},
  year={2017},
  organization={IEEE}
}

@article{sultana2019survey,
  title={Survey on SDN based network intrusion detection system using machine learning approaches},
  author={Sultana, Nasrin and Chilamkurti, Naveen and Peng, Wei and Alhadad, Rabei},
  journal={Peer-to-Peer Networking and Applications},
  volume={12},
  number={2},
  pages={493--501},
  year={2019},
  publisher={Springer}
}

@article{alshamrani2019survey,
  title={A survey on advanced persistent threats: Techniques, solutions, challenges, and research opportunities},
  author={Alshamrani, Adel and Myneni, Sowmya and Chowdhary, Ankur and Huang, Dijiang},
  journal={IEEE Communications Surveys \& Tutorials},
  volume={21},
  number={2},
  pages={1851--1877},
  year={2019},
  publisher={IEEE}
}

@article{elsayed2020insdn,
  title={InSDN: A novel SDN intrusion dataset},
  author={Elsayed, Mahmoud Said and Le-Khac, Nhien-An and Jurcut, Anca D},
  journal={IEEE Access},
  volume={8},
  pages={165263--165284},
  year={2020},
  publisher={IEEE}
}

@article{lu2018learning,
  title={Learning under concept drift: A review},
  author={Lu, Jie and Liu, Anjin and Dong, Fan and Gu, Feng and Gama, Joao and Zhang, Guangquan},
  journal={IEEE Transactions on Knowledge and Data Engineering},
  volume={31},
  number={12},
  pages={2346--2363},
  year={2018},
  publisher={IEEE}
}

@inproceedings{abubakar2017machine,
  title={Machine learning based intrusion detection system for software defined networks},
  author={Abubakar, Atiku and Pranggono, Bernardi},
  booktitle={2017 seventh international conference on emerging security technologies (EST)},
  pages={138--143},
  year={2017},
  organization={IEEE}
  }

@article{gama2014survey,
  title={A survey on concept drift adaptation},
  author={Gama, Jo{\~a}o and {\v{Z}}liobait{\.e}, Indr{\.e} and Bifet, Albert and Pechenizkiy, Mykola and Bouchachia, Abdelhamid},
  journal={ACM computing surveys (CSUR)},
  volume={46},
  number={4},
  pages={1--37},
  year={2014},
  publisher={ACM New York, NY, USA}
}

@article{gomes2017adaptive,
  title={Adaptive random forests for evolving data stream classification},
  author={Gomes, Heitor M and Bifet, Albert and Read, Jesse and Barddal, Jean Paul and Enembreck, Fabr{\'\i}cio and Pfharinger, Bernhard and Holmes, Geoff and Abdessalem, Talel},
  journal={Machine Learning},
  volume={106},
  number={9},
  pages={1469--1495},
  year={2017},
  publisher={Springer}
}

@article{halabi2021protecting,
  title={Protecting the Internet of vehicles against advanced persistent threats: a bayesian Stackelberg game},
  author={Halabi, Talal and Wahab, Omar Abdel and Al Mallah, Ranwa and Zulkernine, Mohammad},
  journal={IEEE Transactions on Reliability},
  volume={70},
  number={3},
  pages={970--985},
  year={2021},
  publisher={IEEE}
}

@article{raab2020reactive,
  title={Reactive soft prototype computing for concept drift streams},
  author={Raab, Christoph and Heusinger, Moritz and Schleif, Frank-Michael},
  journal={Neurocomputing},
  volume={416},
  pages={340--351},
  year={2020},
  publisher={Elsevier}
}

@article{frias2014online,
  title={Online and non-parametric drift detection methods based on Hoeffding’s bounds},
  author={Frias-Blanco, Isvani and del Campo-{\'A}vila, Jos{\'e} and Ramos-Jimenez, Gonzalo and Morales-Bueno, Rafael and Ortiz-Diaz, Agustin and Caballero-Mota, Yaile},
  journal={IEEE Transactions on Knowledge and Data Engineering},
  volume={27},
  number={3},
  pages={810--823},
  year={2014},
  publisher={IEEE}
}

@inproceedings{bifet2007learning,
  title={Learning from time-changing data with adaptive windowing},
  author={Bifet, Albert and Gavalda, Ricard},
  booktitle={Proceedings of the 2007 SIAM international conference on data mining},
  pages={443--448},
  year={2007},
  organization={SIAM}
}

@inproceedings{gama2004learning,
  title={Learning with drift detection},
  author={Gama, Joao and Medas, Pedro and Castillo, Gladys and Rodrigues, Pedro},
  booktitle={Brazilian symposium on artificial intelligence},
  pages={286--295},
  year={2004},
  organization={Springer}
}

@inproceedings{baena2006early,
  title={Early drift detection method},
  author={Baena-Garc{\i}a, Manuel and del Campo-{\'A}vila, Jos{\'e} and Fidalgo, Ra{\'u}l and Bifet, Albert and Gavalda, R and Morales-Bueno, Rafael},
  booktitle={Fourth international workshop on knowledge discovery from data streams},
  volume={6},
  pages={77--86},
  year={2006}
}

@article{chandola2009anomaly,
  title={Anomaly detection: A survey},
  author={Chandola, Varun and Banerjee, Arindam and Kumar, Vipin},
  journal={ACM computing surveys (CSUR)},
  volume={41},
  number={3},
  pages={1--58},
  year={2009},
  publisher={ACM New York, NY, USA}
}

@incollection{mulimani2021adaptive,
  title={Adaptive Ensemble Learning with Concept Drift Detection for Intrusion Detection},
  author={Mulimani, Deepa and Totad, Shashikumar G and Patil, Prakashgoud and Seeri, Shivananda V},
  booktitle={Data Engineering and Intelligent Computing},
  pages={331--339},
  year={2021},
  publisher={Springer}
}

@inproceedings{yuan2018concept,
  title={A concept drift based ensemble incremental learning approach for intrusion detection},
  author={Yuan, Xiaoming and Wang, Ran and Zhuang, Yi and Zhu, Kun and Hao, Jie},
  booktitle={2018 IEEE international conference on internet of things (IThings) and IEEE green computing and communications (GreenCom) and IEEE cyber, physical and social computing (CPSCom) and IEEE smart data (SmartData)},
  pages={350--357},
  year={2018},
  organization={IEEE}
}

@inproceedings{dasu2006information,
  title={An information-theoretic approach to detecting changes in multi-dimensional data streams},
  author={Dasu, Tamraparni and Krishnan, Shankar and Venkatasubramanian, Suresh and Yi, Ke},
  booktitle={In Proc. Symp. on the Interface of Statistics, Computing Science, and Applications},
  year={2006},
  organization={Citeseer}
}

@inproceedings{andresini2021network,
  title={A Network Intrusion Detection System for Concept Drifting Network Traffic Data},
  author={Andresini, Giuseppina and Appice, Annalisa and Loglisci, Corrado and Belvedere, Vincenzo and Redavid, Domenico and Malerba, Donato},
  booktitle={International Conference on Discovery Science},
  pages={111--121},
  year={2021},
  organization={Springer}
}

@inproceedings{horchulhack2022stream,
  title={A Stream Learning Intrusion Detection System for Concept Drifting Network Traffic},
  author={Horchulhack, Pedro and Viegas, Eduardo K and Lopez, Martin Andreoni},
  booktitle={2022 6th Cyber Security in Networking Conference (CSNet)},
  pages={1--7},
  year={2022},
  organization={IEEE}
}

@article{priya2021deep,
  title={Deep learning framework for handling concept drift and class imbalanced complex decision-making on streaming data},
  author={Priya, S and Uthra, R Annie},
  journal={Complex \& Intelligent Systems},
  pages={1--17},
  year={2021},
  publisher={Springer}
}

@article{martindale2020ensemble,
  title={Ensemble-based online machine learning algorithms for network intrusion detection systems using streaming data},
  author={Martindale, Nathan and Ismail, Muhammad and Talbert, Douglas A},
  journal={Information},
  volume={11},
  number={6},
  pages={315},
  year={2020},
  publisher={MDPI}
}

@article{yang2021lightweight,
  title={A lightweight concept drift detection and adaptation framework for IoT data streams},
  author={Yang, Li and Shami, Abdallah},
  journal={IEEE Internet of Things Magazine},
  volume={4},
  number={2},
  pages={96--101},
  year={2021},
  publisher={IEEE}
}

@inproceedings{myneni2020dapt,
  title={DAPT 2020-constructing a benchmark dataset for advanced persistent threats},
  author={Myneni, Sowmya and Chowdhary, Ankur and Sabur, Abdulhakim and Sengupta, Sailik and Agrawal, Garima and Huang, Dijiang and Kang, Myong},
  booktitle={International Workshop on Deployable Machine Learning for Security Defense},
  pages={138--163},
  year={2020},
  organization={Springer}
}

@misc{RFfeature, title = {Feature importances with a forest of trees}, url={https://scikit-learn.org/stable/auto_examples/ensemble/plot_forest_importances.html}, journal={RandomForestClassifier}}

@misc{StandardScaler, title = {StandardScaler}, url={https://scikit-learn.org/stable/modules/generated/sklearn.preprocessing.StandardScaler.html}, journal={StandardScaler}}

@inproceedings{ullah2020scheme,
  title={A scheme for generating a dataset for anomalous activity detection in iot networks},
  author={Ullah, Imtiaz and Mahmoud, Qusay H},
  booktitle={Advances in Artificial Intelligence: 33rd Canadian Conference on Artificial Intelligence, Canadian AI 2020, Ottawa, ON, Canada, May 13--15, 2020, Proceedings 33},
  pages={508--520},
  year={2020},
  organization={Springer}
}

@inproceedings{dos2021improving,
  title={Improving intrusion detection confidence through a moving target defense strategy},
  author={dos Santos, Roger R and Viegas, Eduardo K and Santin, Altair O},
  booktitle={2021 IEEE Global Communications Conference (GLOBECOM)},
  pages={1--6},
  year={2021},
  organization={IEEE}
}

@article{alqahtanidetecting,
  title={Detecting Stealthy Scans in SDN using a Hybrid Intrusion Detection System},
  author={Alqahtani, Abdullah H and Clark, John A},
  booktitle={WRIT – Workshop on Research for Insider Threats},
  year={2022},
  organization={ACSAC 2022}
}

@misc{OnLineSGDOneClassSVM, 
  title={Stochastic Gradient Descent},
  url={https://scikit-learn.org/stable/modules/sgd.html},
  journal={SGDOneClassSVM}
}

@article{girdler2021implementing,
  title={Implementing an intrusion detection and prevention system using Software-Defined Networking: Defending against ARP spoofing attacks and Blacklisted MAC Addresses},
  author={Girdler, Thomas and Vassilakis, Vassilios G},
  journal={Computers \& Electrical Engineering},
  volume={90},
  pages={106990},
  year={2021},
  publisher={Elsevier}
}

@article{lu2016concept,
  title={A concept drift-tolerant case-base editing technique},
  author={Lu, Ning and Lu, Jie and Zhang, Guangquan and De Mantaras, Ramon Lopez},
  journal={Artificial Intelligence},
  volume={230},
  pages={108--133},
  year={2016},
  publisher={Elsevier}
}

@article{lu2014concept,
  title={Concept drift detection via competence models},
  author={Lu, Ning and Zhang, Guangquan and Lu, Jie},
  journal={Artificial Intelligence},
  volume={209},
  pages={11--28},
  year={2014},
  publisher={Elsevier}
}

@inproceedings{shao2014prototype,
  title={Prototype-based learning on concept-drifting data streams},
  author={Shao, Junming and Ahmadi, Zahra and Kramer, Stefan},
  booktitle={Proceedings of the 20th ACM SIGKDD international conference on Knowledge discovery and data mining},
  pages={412--421},
  year={2014}
}

@inproceedings{kifer2004detecting,
  title={Detecting change in data streams},
  author={Kifer, Daniel and Ben-David, Shai and Gehrke, Johannes},
  booktitle={VLDB},
  volume={4},
  pages={180--191},
  year={2004},
  organization={Toronto, Canada}
}

@inproceedings{song2007statistical,
  title={Statistical change detection for multi-dimensional data},
  author={Song, Xiuyao and Wu, Mingxi and Jermaine, Christopher and Ranka, Sanjay},
  booktitle={Proceedings of the 13th ACM SIGKDD international conference on Knowledge discovery and data mining},
  pages={667--676},
  year={2007}
}

@inproceedings{qahtan2015pca,
  title={A pca-based change detection framework for multidimensional data streams: Change detection in multidimensional data streams},
  author={Qahtan, Abdulhakim A and Alharbi, Basma and Wang, Suojin and Zhang, Xiangliang},
  booktitle={Proceedings of the 21th ACM SIGKDD International Conference on Knowledge Discovery and Data Mining},
  pages={935--944},
  year={2015}
}

@inproceedings{gu2016concept,
  title={Concept drift detection based on equal density estimation},
  author={Gu, Feng and Zhang, Guangquan and Lu, Jie and Lin, Chin-Teng},
  booktitle={2016 International Joint Conference on Neural Networks (IJCNN)},
  pages={24--30},
  year={2016},
  organization={IEEE}
}

@article{bu2017incremental,
  title={An incremental change detection test based on density difference estimation},
  author={Bu, Li and Zhao, Dongbin and Alippi, Cesare},
  journal={IEEE Transactions on Systems, Man, and Cybernetics: Systems},
  volume={47},
  number={10},
  pages={2714--2726},
  year={2017},
  publisher={IEEE}
}

@inproceedings{liu2017regional,
  title={Regional concept drift detection and density synchronized drift adaptation},
  author={Liu, Anjin and Song, Yiliao and Zhang, Guangquan and Lu, Jie},
  booktitle={IJCAI International Joint Conference on Artificial Intelligence},
  year={2017}
}

@article{bu2016pdf,
  title={A pdf-free change detection test based on density difference estimation},
  author={Bu, Li and Alippi, Cesare and Zhao, Dongbin},
  journal={IEEE transactions on neural networks and learning systems},
  volume={29},
  number={2},
  pages={324--334},
  year={2016},
  publisher={IEEE}
}

@article{breiman2001random,
  title={Random forests},
  author={Breiman, Leo},
  journal={Machine learning},
  volume={45},
  pages={5--32},
  year={2001},
  publisher={Springer}
}

@inproceedings{alqahtani2022enhanced,
  title={Enhanced Scanning in SDN Networks and its Detection using Machine Learning},
  author={Alqahtani, Abdullah H and Clark, John A},
  booktitle={2022 IEEE 4th International Conference on Trust, Privacy and Security in Intelligent Systems, and Applications (TPS-ISA)},
  pages={188--197},
  year={2022},
  organization={IEEE}
}

@article{robbins1951stochastic,
  title={A stochastic approximation method},
  author={Robbins, Herbert and Monro, Sutton},
  journal={The annals of mathematical statistics},
  pages={400--407},
  year={1951},
  publisher={JSTOR}
}

\begin{subappendices}
\renewcommand{\thesection}{\Alph{section}}%

\section{}
\label{appendexA}
Details of the scenario of attacks and normal traffic in DAPT 2020 dataset are shown in Table \ref{table: DAPT2020discription}.

\begin{table*}
\caption{DAPT 2020 dataset attacks scenario}
\centering
\setlength\tabcolsep{2.5pt}
\begin{tabular}{|c | c |c |p{4cm} |c|} 
\hline 
\textbf{day} & \textbf{Benign} & \textbf{attack stage}  & \textbf{attack type} &\textbf{attack time}  \\ 
\hline
1 & $\checkmark$ & $\times$   & - &  - \\
\hline
2 & \checkmark & Reconnaissance & scanning & 8:00 a.m. - 6:00 p.m. \\
\hline
3 & $\checkmark$ & Foothold establishment  &  Sqli, XSS, authentication bypass &  8:00 a.m. - 6:00 p.m. \\
\hline
4 & $\checkmark$ & Lateral Movement & network scan (insider), authentication bypass, SQLi &  8:00 a.m. - 6:00 p.m. \\
\hline
5 & $\checkmark$ &   Data exfiltration & Exfiltrating data to C\&C & 8:00 a.m. - 6:00 p.m. \\
\hline
\end{tabular}
\label{table: DAPT2020discription}
\end{table*}

\section{}
\label{appendexB}
Table \ref{table:InSDN} shows details of the attack scenario and normal traffic in InSDN dataset.

\begin{table*}
\caption{Attacks details on InSDN dataset}
\centering
\begin{tabular}{|c | c |c |c|} 
\hline 
\textbf{day} & \textbf{Data group} & \textbf{attack type} &\textbf{attack time}  \\ 
\hline
1 & OVS  &  Web attack & 25/12/2019 (17:00 - 23:00) \\
\hline
2 & OVS & Probe & 26/12/2019 ( 13:00 - 15:00) \\
\hline
3 & OVS & DoS & 27/12/2019 (16:00 - 19:00) \\
\hline
4 & OVS & Botnet & 31/12/2019 (14:00 - 15:00) \\
  & OVS & DDoS & 31/12/2019 (18:00 - 19:00) \\
  & Normal &   - & 31/12/2019 (14:00 - 15:00) \\
\hline
5 & OVS & DoS & 01/01/2020 (22:00 - 24:00) \\
\hline
6 & Metasploitable-2 & DoS & 09/01/2020 (19:00 - 20:00) \\
  & Metasploitable-2 & Probe & 09/01/2020 (16:00 - 18:00) \\
  & Metasploitable-2 & U2R & 09/01/2020 (21:00 - 22:00) \\
\hline
7 & Metasploitable-2 & BFA & 10/01/2020 (01:00 - 04:00) \\
  & Metasploitable-2 & DDoS & 10/01/2020 (05:00 - 06:00) \\
  & Metasploitable-2 & U2R & 10/01/2020 (00:00 - 21:00) \\
\hline
8 & OVS & BFA & 12/01/2020 (01:00 - 02:00) \\
  & OVS & DDoS & 12/01/2020 (16:00 - 16:00) \\
\hline
9 & Normal & - & 04/02/2020 ( 12:00 - 13:00) \\
  & Normal & - & 04/02/2020 ( 12:00 - 13:00) \\
\hline
10 & Normal & - & 05/02/2020 ( 11:00 - 20:00) \\
\hline
11 & Normal & - & 07/02/2020 ( 13:00 - 20:00) \\

\hline
\end{tabular}
\label{table:InSDN}
\end{table*}

\section{}
\label{appendexC}
In this section, Table \ref{table:CICIDS2017} shows the normal and attack details of CICIDS 2017 dataset.

\begin{table*}
\caption{Attacks details on CICIDS 2017 dataset}
\centering
\begin{tabular}{|c | c |c |c |c|} 
\hline 
\textbf{day} & \textbf{Benign} & \textbf{attack category}  & \textbf{attack type} &\textbf{attack time}  \\ 
\hline
1 & $\checkmark$ & $\times$   & - &  - \\
\hline
2 & $\checkmark$ & Brute Force & FTP-Patator & 9:20 – 10:20 a.m. \\
  &                &             &  SSH-Patator & 14:00 – 15:00 p.m.  \\   
\hline
3 & $\checkmark$ & DoS / DDoS  &  DoS slowloris & 9:47 – 10:10 a.m. \\
   &              &             & DoS Slowhttptest & 10:14 – 10:35 a.m. \\
   &              &             & DoS Hulk & 10:43 – 11 a.m. \\
   &              &             & DoS GoldenEye & 11:10 – 11:23 a.m. \\
   &              & Heartbleed  & Heartbleed Port 444 & 15:12 - 15:32 \\
\hline
4 & $\checkmark$ & Web Attack & Brute Force & 9:20 – 10 a.m. \\
  &              &             & XSS & 10:15 – 10:35 a.m. \\
  &              &             & Sql Injection & 10:40 – 10:42 a.m. \\
  &              & Infiltration – Dropbox download & Meta exploit Win Vista & 14:19 \& 14:20-14:21 p.m. \\
  &              &                                 &                          & 14:33 -14:35 \\
  &              & Infiltration – Cool disk  & MAC & 14:53 p.m. – 15:00 p.m \\
  &              & Infiltration – Dropbox download & Win Vista & 15:04 – 15:45 p.m. \\  
\hline
5 & $\checkmark$ &   Botnet ARES & & 10:02 a.m. – 11:02 a.m. \\
  &              &    probe      & port scan & 13:55 - 14:35\\
  &              &               &           & 14:51 - 15:29 \\
  &              &    DDoS       & LOIT      & 15:56 – 16:16 \\
\hline
\end{tabular}
\label{table:CICIDS2017}
\end{table*}

\end{subappendices}

\end{document}